# Sparse Multiband Signal Acquisition Receiver with Co-prime Sampling

Yijiu Zhao, *Member, IEEE*, and Shuangman Xiao

*Abstract*—Cognitive radio (CR) requires spectrum sensing over a broad frequency band. One of the crucial tasks in CR is to sample wideband signal at high sampling rate. In this paper, we propose an acquisition receiver with co-prime sampling technique for wideband sparse signals, which occupy a small part of band range. In this proposed acquisition receiver, we use two low speed analog-to-digital converters (ADCs) to capture a common sparse multiband signal, whose band locations are unknown. The two ADCs are synchronously clocked at co-prime sampling rates. The obtained samples are re-sequenced into a group of uniform sequences with low rate. We derive the mathematical model for the receiver in the frequency domain and present its signal reconstruction algorithm. Compared to the existing sub-Nyquist sampling techniques, such as multi-coset sampling and modulated wideband converter, the proposed approach has a simple system architecture and can be implemented with only two samplers. Experimental results are reported to demonstrate the feasibility and advantage of the proposed model. For sparse multiband signal with unknown spectral support, the proposed system requires a sampling rate much lower than Nyquist rate, while produces satisfactory reconstruction.

*Index Terms*—Co-prime sampling, cognitive radio, multi-coset sampling, compressed sensing, spectrum sensing.

## I. INTRODUCTION

IN wireless communication, wide spectral band is divided into narrowband slices. Before being transmitted in those spectral slices, baseband signal is modulated by high carrier frequency. The increasing demand from new wireless communication users has become a crucial issue recently. However, studies [1], [2] have shown that this over-crowded spectrum is usually underutilized. Although cognitive radio (CR) would allow secondary users to use the licensed spectral slices when the corresponding primary users are not active [3], CR typically monitors the spectrum based on the Nyquist sampling theorem and requires high sampling rate. The Nyquist rates of signals are high and may exceed the specifications of best commercial analog-to-digital converters (ADCs). ADC has become a bottleneck of high speed acquisition system.

In CR, signals can be described as the union of a small number of narrowband transmissions over a wide band range. Consequently, signal spectrum is often sparse and can be modeled as sparse multiband signal. For such sparse multiband signal, it may be reconstructed from sub-Nyquist samples in a more intelligent way.

Multi-coset sampling (MCS) is a non-uniform periodic sub-Nyquist sampling method [4], [5] for sparse multiband signal. MCS consists of a bank of ADCs clocked at the same rate but with different phase delays to facilitate concurrent sampling of the wideband signal. The performance of MCS hinges upon the accurate clock phase delays, which is complex and costly. On the other hand, the number of ADCs is proportional to the number of transmissions. Compared to MCS, multi-rate sampling (MRS) [6] needs less channels. Most of MRS's channels could operate at sub-Nyquist rate, while the sampled spectrum should be unaliased in at least one of the channels. Co-prime sampling [7], [8] was proposed to estimate directions-of-arrival (DOA) in the spatial domain. In the DOA estimation, co-prime array can increase the degree of freedom and improve the estimation accuracy. In [9], a sampling model with three channels was proposed to estimate the frequencies of multiple sinusoids. The three channels work at sub-Nyquist rate, whose undersampling ratios are pairwise co-prime. It is an efficient approach for harmonic sparse signal sampling. Sparse fast Fourier transform (sFFT) [10], [11] algorithm provides a low bound for number of samples. Specifically, to achieve the bound, the coefficients of signal need to be pseudo-randomly permuted. sFFT requires custom ADCs that can randomly sub-sample the signal with inter-sample spacing as small as the inverse of the signal bandwidth.

Compressed sensing (CS) is an emerging alternative approach for the acquisition of sparse signals [12], [13]. For spectral sparse signal, the idea of CS is that the spectral information is much smaller than its bandwidth, and signal can be reconstructed from its low dimensional samples. Employing random demodulation technique, Kirolos *et al*. [14], [15] developed an analog-to-information converter (AIC) to realize sub-Nyquist rate sampling of wideband signal using CS reconstruction. For wideband harmonic sparse signal, AIC has been shown an effective sampling method. While AIC does not work for sparse multiband signal that is widely used in CR. Eldar *et al*. [16], [17] developed a modulated wideband converter (MWC). In particular, MWC could be treated as multiple AIC samplers concurrently, and its channels is proportional to the number of sparsely allocated narrow bands in the signal spectrum. MWC works for not only harmonic

Manuscript received xx. xx, 2018.
This work is supported by the National Natural Science Foundation of China (Grant No. 61671114).
Y. Zhao, and S. Xiao are with the School of Automation Engineering, University of Electronic Science and Technology of China, Chengdu 611731, China (e-mail: yijiuzhao@uestc.edu.cn; shuangmanxiao@hotmail.com).

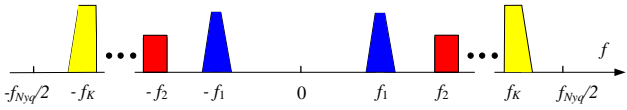

Fig. 1. Spectrum-sparse multiband signal.

sparse signal but also sparse multiband signal. To accomplish sub-Nyquist sampling, both AIC and MWC need to modulate the wideband signal with the pseudorandom pulse sequences at the Nyquist rate. Other CS based sampling techniques are also proposed to improve the sampling rate [18], [19].

Random equivalent sampling (RES) [20], [21] is a non-uniform sampling technique which is widely used in instrumentation applications. RES has a simple architecture and only requires a single ADC clocked at sub-Nyquist rate. In order to produce a successful reconstruction, multiple RES acquisitions need to be run. Moreover, there should be a unique reference point in each RES acquisition, and a time-to-digital converter circuitry should be used to measure the trigger time interval. Compared with uniform sampling techniques, RES takes considerably longer sampling time. Previously [22], [23], CS theory is incorporated into RES technique to enhance the RES signal reconstruction. It has been tested that much fewer samples are required in RES signal reconstruction. Combine random triggering and random demodulation techniques [24], the bandwidth of ADC is avoided. However, RES (or random triggering) based non-uniform sampling techniques require multiple acquisitions, which are not applicable for CR. Non-uniform sampling clock phase also introduces error in reconstructed waveform.

In this work, we propose a sub-Nyquist acquisition receiver based on co-prime sampling technique. Different from conventional co-prime sampling approaches [7-9] (they focus on DOA and harmonic spectral estimation), we consider the sparse multiband signal. In the proposed approach, signals do not need to be pre-processed as random modulated based methods (such as AIC and MWC). In the previous work [25], two co-prime samplers are asynchronously clocked, a time-to-digital converter circuitry is required to measure the phase delay, and it would degrade the performance of system. In this proposed sampling model, a common sparse multiband signal is uniformly and synchronously sampled by only two low speed ADCs (much smaller number of channels than MCS and MRS). These two ADCs are clocked at co-prime sampling rates (hereafter, the proposed model is referred to SCS: synchronous co-prime sampling). After one acquisition, samples are re-sequenced into a group of low speed of sampling sequences. We derive the mathematical expression for SCS, which can be used to blindly reconstruct signal. In comparison to the popular sub-Nyquist sampling approaches, such as e.g. MCS, MWC, RES, the proposed SCS has simple architecture.

The remaining of this paper is organized as follows. In section II, we first introduce the sparse multiband signal model. In Section III We propose synchronous co-prime sampling model and give the comparison to other related works.

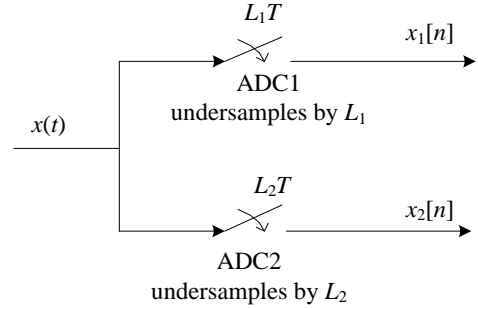

Fig. 2. Scheme of co-prime sampling.

Numerical simulation results are reported and discussed in Section VI, followed by conclusion in Section V.

## II. PROBLEM FORMULATION

A sparse multiband signal is a bandlimited, square-integrable, continuous time signal which consists of a relatively small number of narrow frequency bands over a board spectrum range. Fig. 1 depicts an example of the spectrum of a sparse multiband signal.

Consider a continuous-time sparse multiband signal $x(t)$, which is real-valued and square-integrable, and with the Fourier transform (FT)

$$X(f) = \int_{-\infty}^{+\infty} x(t) e^{-j2\pi ft} dt \ . \quad (1)$$

Denote by $f_{Nyq} = 1/T$ the Nyquist rate of $x(t)$, signal is bandlimited to $\mathcal{F} \subset [-1/(2T), 1/(2T)]$. Let $\mathcal{F}_i$ be the $i^{th}$ active sub-band where $X(f) \neq 0$, whose carrier frequency is $f_i$. The spectral occupation ratio may be defined as $R = \lambda(\mathcal{F})/|[\mathcal{F}]|$ where $\lambda(\mathcal{F})$ is the Lebesgue measure of $\mathcal{F}$, and $|[\mathcal{F}]| = f_{Nyq}$. Since $x(t)$ is spectrum sparse, $R << 1$ should be satisfied. In CR applications, the locations of the signal sub-bands are unknown in advance, signal should be blindly reconstructed.

## III. SUB-NYQUIST CO-PRIME SAMPLING

### A. System Model

Fig. 2 shows an illustrative scheme of SCS. SCS system consists of two sampling channels (ADC1 and ADC2), which uniformly digitize a common input signal. Two ADCs are synchronously clocked at sub-Nyquist rates. Suppose the sampling intervals of two ADCs are $L_1T$ and $L_2T$ respectively. To qualify sub-Nyquist sampling, we require the integer $L_1$ and $L_2$ are much bigger than 1. Moreover, $L_1$ and $L_2$ are required to be relative co-prime integers. Denote by $L$ the least common multiple of $L_1$ and $L_2$, and $L = L_1 \cdot L_2$. The average sampling rate is $(L_1 + L_2)/(LT)$.

Fig. 3 illustrates an example of principle of SCS, where $L_1 = 3$ and $L_2 = 4$ are relative co-prime integers. In Fig. 3, "Nyquist samples" is the digital version of signal with Nyquist rate that is to be reconstructed. Let $x_1[n]$ and $x_2[n]$ denote the samples captured by ADC1 and ADC2. Since ADCs are synchronously clocked, $x_1[n]$ and $x_2[n]$ could be treated as under-samples from Nyquist grids, and they can be expressed as

$$\begin{cases} x_1[n] = x(nL_1T), \\ x_2[n] = x(nL_2T), \end{cases} \quad n \in \mathbb{Z}, \quad (2)$$

where $T$ is the base sampling period.

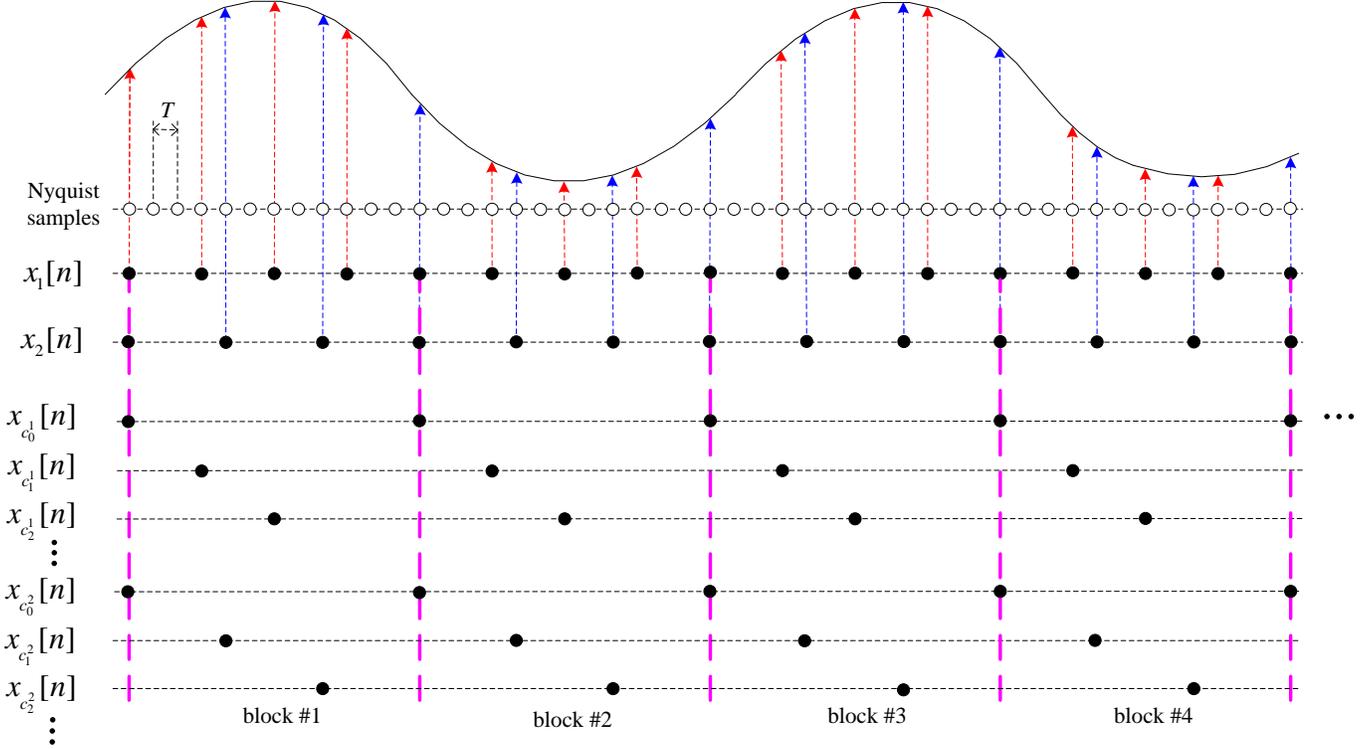

Fig. 3. An illustration of co-prime sampling.

We group $L$ consecutive Nyquist grids in a block. Denote by $L_d$ the greatest common divisor of $L_1$ and $L_2$. There are $L_d$ identical sample (samples) from ADC1 and ADC2 in each block. Since the sampling rates of SCS are co-prime, $L_d = 1$, and there is only one identical sample in each block. If $L_1$ and $L_2$ are not co-prime, $L_d$ would be greater than 1. As a result, SCS generates more valid samples with low average sampling rate.

In each block, denote by $C_1$ and $C_2$ the time stamps that describe the indexes of samples obtained by ADC1 and ADC2 respectively. The sets $C_1$ and $C_2$ are referred to as the sampling patterns where

$$C_1 = \{c_i^1 \mid c_i^1 = L_1 i : i = 0, 1, \ldots, L_2 - 1\}, \quad (3)$$

and

$$C_2 = \{c_i^2 \mid c_i^2 = L_2 i : i = 0, 1, \ldots, L_1 - 1\}. \quad (4)$$

Obviously, SCS can obtains $M$ ($M = L_1 + L_2 - 1$) different non-uniform samples in each block, and their indexes are described by sampling pattern $C = C_1 \cup C_2$. We could construct $M$ uniform sampling sequences, which pick one sample from each block, and the index is given by $C$.

There are $L_2$ samples in each block from ADC1, and then $L_2$ uniform sampling sequences can be constructed

$$x_{c_i^1}[n] = \begin{cases} x(nLT + c_i^1 T), & n \in \mathbb{Z} \\ 0, & \text{otherwise} \end{cases}, \quad (5)$$

and $L_1$ sampling sequences constructed from samples obtained by ADC2

$$x_{c_i^2}[n] = \begin{cases} x(nLT + c_i^2 T), & n \in \mathbb{Z} \\ 0, & \text{otherwise} \end{cases}. \quad (6)$$

For sampling sequence $x_{c_i^1}[n]$, its discrete-time Fourier transform (DTFT) $X_{c_i^1}\left(e^{j2\pi fLT}\right)$ can be directly calculated:

$$\begin{aligned} X_{c_i^1}\left(e^{j2\pi fLT}\right) &= \sum_{n=-\infty}^{+\infty} x_{c_i^1}[n]\exp(-j2\pi fnLT) \\ &= \sum_{n=-\infty}^{+\infty} x(nLT + c_i^1 T)\exp(-j2\pi fnLT) \\ &= \frac{1}{LT}\sum_{l=-L/2+1}^{L/2} X\left(f + \frac{l}{LT}\right)\exp\left(j2\pi c_i^1 T\left(f + \frac{l}{LT}\right)\right) \end{aligned}, \quad (7)$$

where $L$ is assumed to be an even number and $i \in \{0, 1, \ldots, L_2 -$

1}. Because $x(t)$ is assumed to be bandlimited to $\mathcal{F}$, we have finite summation limits in the last equation in (7). Since $X_{c_i^1}\left(e^{j2\pi fLT}\right)$ is periodic with period $1/(LT)$, we can choose only one period of $X_{c_i^1}\left(e^{j2\pi fLT}\right)$. Here we restrict $f \in \mathcal{F}_0$, and

$$\mathcal{F}_0 = \left[0, \frac{1}{LT}\right). \tag{8}$$

For odd $L$, in (7), the range of $l$ should be $[-(L-1)/2, (L-1)/2]$, and $\mathcal{F}_0 = [-1/(2LT), 1/(2LT)]$.

Define $Y_{c_i^1}\left(e^{j2\pi fLT}\right)$ as:

$$\begin{aligned}Y_{c_i^1}\left(e^{j2\pi fLT}\right) &= \exp\left(-j2\pi f c_i^1 T\right) X_{c_i^1}\left(e^{j2\pi fLT}\right) \\ &= \frac{1}{LT} \sum_{l=-L/2+1}^{L/2} X\left(f + \frac{l}{LT}\right) \exp\left(j2\pi c_i^1 \frac{l}{L}\right)\end{aligned}, \tag{9}$$

where $f \in \mathcal{F}_0$. Denote by $y_{c_i^1}[n]$ the inverse DTFT of $Y_{c_i^1}\left(e^{j2\pi fLT}\right)$. Clearly, $y_{c_i^1}[n]$ is the shifted version of $x_{c_i^1}[n]$, which is right shifted by $c_i^1 T$.

Similarly, we define sequence $y_{c_i^2}[n]$, and its DTFT can be expressed as:

$$\begin{aligned}Y_{c_i^2}\left(e^{j2\pi fLT}\right) &= \exp\left(-j2\pi f c_i^2 T\right) X_{c_i^2}\left(e^{j2\pi fLT}\right) \\ &= \frac{1}{LT} \sum_{l=-L/2+1}^{L/2} X\left(f + \frac{l}{LT}\right) \exp\left(j2\pi c_i^2 \frac{l}{L}\right)\end{aligned}, \tag{10}$$

where $f \in \mathcal{F}_0$. Since samples indexed by $c_0^1$ and $c_0^2$ are identical, we require $i \in \{1, ..., L_1 - 1\}$ in (10).

We arrange $Y_{c_i^1}\left(e^{j2\pi fLT}\right)$ and $Y_{c_i^2}\left(e^{j2\pi fLT}\right)$ in a vector $\mathbf{y}(f)$. Denote by $\mathbf{y}_i(f)$ the $i^{th}$ entry of $\mathbf{y}(f)$, and it can be written as:

$$\mathbf{y}_i(f) = \begin{cases} Y_{c_i^1}\left(e^{j2\pi fLT}\right) & 0 \leq i < L_2 \\ Y_{c_{i-L_2+1}^2}\left(e^{j2\pi fLT}\right) & L_2 \leq i < M \end{cases}. \tag{11}$$

Define matrix $\mathbf{\Phi}$ as:

$$\mathbf{\Phi}_{i,l} = \begin{cases} \dfrac{1}{LT}\exp\left(j2\pi c_i^1 \dfrac{K_l}{L}\right) & 0 \leq i < L_2 \\ \dfrac{1}{LT}\exp\left(j2\pi c_{i-L_2+1}^2 \dfrac{K_l}{L}\right) & L_2 \leq i < M \end{cases}, \tag{12}$$

where $l = 0, ..., L - 1$, and $K_l = -L/2 + 1 + l$.

Combine (9) and (10), we have vector-matrix form as

$$\mathbf{y}(f) = \mathbf{\Phi}\mathbf{x}(f) \quad f \in \mathcal{F}_0, \tag{13}$$

and the $l^{th}$ entry of $\mathbf{x}(f)$ has the following expression

$$\mathbf{x}_l(f) = X\left(f + \frac{K_l}{LT}\right). \tag{14}$$

Obviously, $X(f)$ is partitioned into $L$ spectral slices with equal width of $1/(LT)$ that are the entries of $\mathbf{x}(f)$. According to (9) and (10), $\mathbf{y}_i(f)$ is also given by

$$\mathbf{y}_i(f) = \begin{cases} \exp\left(-j2\pi f c_i^1 T\right) X_{c_i^1}\left(e^{j2\pi fLT}\right) & 0 \leq i < L_2 \\ \exp\left(-j2\pi f c_{i-L_2+1}^2 T\right) X_{c_{i-L_2+1}^2}\left(e^{j2\pi fLT}\right) & L_2 \leq i < M \end{cases}. \tag{15}$$

Although, the above analysis is about the real-valued signal, the results are also applicable to complex-valued signals.

*B. Signal Reconstruction*

Eqn. (13) ties the FT of unknown signal and sampling sequences in the frequency domain. To reconstruct signal $x(t)$, we must first solve problem (13). However, $M < L$, (13) is an under determined problem. Fortunately, $\mathbf{x}(f)$ is sparse, and most of entries of $\mathbf{x}(f)$ are zeros. The task of *support estimation* is to identify which entries of $\mathbf{x}(f)$ contain active bands. Support estimation could be realized by examining the covariance matrix of the SCS sequences. And the support could be teased out using multiple signal classification (MUSIC) algorithm [26].

Let $S$ denote the index set that marks the locations of the nonzero entries of $\mathbf{x}(f)$. $\mathbf{x}(f)$ is $|S|$-sparse. In order to solve (13), we require that $|S| < M$. Support estimation means to find $S$, and problem (13) can be rewritten as

$$\mathbf{y}(f) = \mathbf{\Phi}_S \mathbf{x}^S(f), \tag{16}$$

where $\mathbf{\Phi}_S$ denotes submatrix composed of the columns of $\mathbf{\Phi}$ indexed by $S$, and vector $\mathbf{x}^S(f)$ contains the non-zero entries of $\mathbf{x}(f)$ indexed by $S$.

Consider the $M \times M$ covariance matrix of $\mathbf{y}(f)$,

$$\begin{aligned}\mathbf{R_y} &= \int_{f \in \mathcal{F}_0} \mathbf{y}(f)\mathbf{y}(f)^H \, df \\ &= \mathbf{\Phi}\left[\int_{f \in \mathcal{F}_0} \mathbf{x}(f)\mathbf{x}(f)^H \, df\right]\mathbf{\Phi}^H \\ &= \mathbf{\Phi}\mathbf{R_x}\mathbf{\Phi}^H\end{aligned}, \tag{17}$$

where $\mathbf{R_x}$ is the covariance matrix of $\mathbf{x}(f)$, and $H$ denotes Hermitian transpose. Since $\mathbf{x}(f)$ is $|S|$-sparse, only $|S|$ of columns and rows of $\mathbf{R_x}$ will be nonzero, and $\mathbf{R_y}$ can be reduced as

$$\mathbf{R_y} = \mathbf{\Phi}_S \left[\mathbf{R_x}\right]_S \left[\mathbf{\Phi}_S\right]^H. \tag{18}$$

Note from (12), $\mathbf{\Phi}$ has full rank. Consequently, its sub-matrix $\mathbf{\Phi}_S$ also has full rank, and rank($\mathbf{\Phi}_S$) = $|S|$ ($M > |S|$ is assumed). If rank($[\mathbf{R_x}]_S$) = $|S|$, then (18) implies rank($\mathbf{R_y}$) = $|S|$ (proof is given in Appendix A).

Consider the eigen-decomposition of $\mathbf{R_y}$. Let $\mathbf{\Sigma}_r$ and $\mathbf{\Sigma}_n$ denote the diagonal matrices containing $|S|$ nonzero and $M - |S|$ zero eigenvalues respectively, and $\mathbf{U}_r$ and $\mathbf{U}_n$ are associated eigenvectors matrices. The eigen-decomposition may be expressed as

$$\mathbf{R_y} = [\mathbf{U}_r, \mathbf{U}_n] \begin{bmatrix} \mathbf{\Sigma}_r & O \\ O & \mathbf{\Sigma}_n \end{bmatrix} \begin{bmatrix} [\mathbf{U}_r]^H \\ [\mathbf{U}_n]^H \end{bmatrix}. \quad (19)$$
$$= \mathbf{U}_r \mathbf{\Sigma}_r [\mathbf{U}_r]^H$$

The columns of $\mathbf{U}_r$ span the same space as that of the columns of $\mathbf{R_y}$, and range($\mathbf{R_y}$) = range($\mathbf{U}_r$). Similarly, (18) implies range($\mathbf{R_y}$) = range($\mathbf{\Phi}_S$), and we have range($\mathbf{U}_r$) = range($\mathbf{\Phi}_S$). Therefore, the columns of $\mathbf{\Phi}$ indexed by $S$ should lie in the space spanned by the columns of $\mathbf{U}_r$.

Denote by $\mathbf{P}_{\mathbf{U}_r}$ the projection matrix of the space spanned by the columns of $\mathbf{U}_r$. $\mathbf{P}_{\mathbf{U}_r} := \mathbf{U}_r [\mathbf{U}_r]^H$. If the column index $i \in S$, then the projection $\mathbf{P}_{\mathbf{U}_r} \mathbf{\Phi}_i$ will have significant $l_2$ norm,

$$\left\| \mathbf{P}_{\mathbf{U}_r} \mathbf{\Phi}_i \Big|_{i \in S} \right\|_2 > \left\| \mathbf{P}_{\mathbf{U}_r} \mathbf{\Phi}_k \Big|_{k \notin S} \right\|_2, \quad (20)$$

and

$$[\mathbf{\Phi}_i]^H \mathbf{U}_r [\mathbf{U}_r]^H \mathbf{\Phi}_i \Big|_{i \in S} > [\mathbf{\Phi}_k]^H \mathbf{U}_r [\mathbf{U}_r]^H \mathbf{\Phi}_k \Big|_{k \notin S}. \quad (21)$$

Based on (21), the support of $\mathbf{x}(f)$ can be estimated.

Because $|S| < M$ is assumed, (16) is an over determined problem. Once the support is obtained, (16) can be solved,

$$\mathbf{x}^S(f) = [\mathbf{\Phi}_S]^\dagger \mathbf{y}(f)$$
$$= \left( [\mathbf{\Phi}_S]^H \mathbf{\Phi}_S \right)^{-1} [\mathbf{\Phi}_S]^H \mathbf{y}(f), \quad (22)$$

where $\dagger$ denotes pseudoinverse.

### C. Comparison to the Related Works

*Multi-coset sampling (MCS)* [27]: MCS is a selection of certain samples from Nyquist grids. Denote by $x(nT)$ the sampling sequence on Nyquist grids. Let $P$ be a positive integer, and the set $D = \{d_i\}_{i=0}^{p-1}$ be the sampling pattern with $0 \leq d_0 \leq \ldots d_{p-1} \leq P - 1$. MCS consists of $p$ uniform sequences, with the $i^{\text{th}}$ sequence defined by

$$x_{d_i}[n] = x(nPT + d_iT), \quad n \in \mathbb{Z}. \quad (23)$$

In reconstruction stage, $p$ sampling sequences are used. Therefore, the average sampling rate is $p/(PT)$. Generally, $p \ll P$, the average sampling rate is much lower than Nyquist rate.

Fig. 4 depicts the scheme of MCS. It consists of $p$ samplers, and $p$ is proportional to number of active signal sub-bands. The large number of sampling channels makes MCS difficult to be implemented.

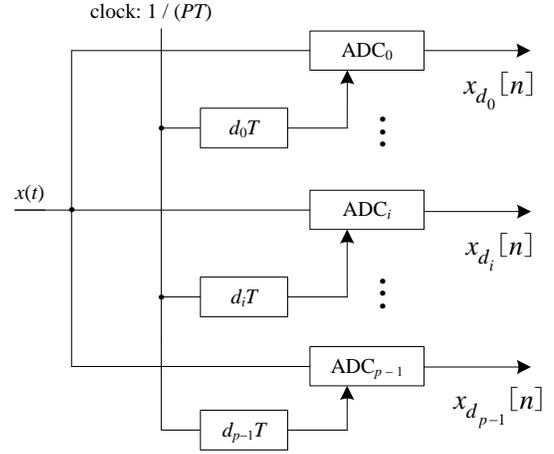

Fig. 4. Block diagram of MCS.

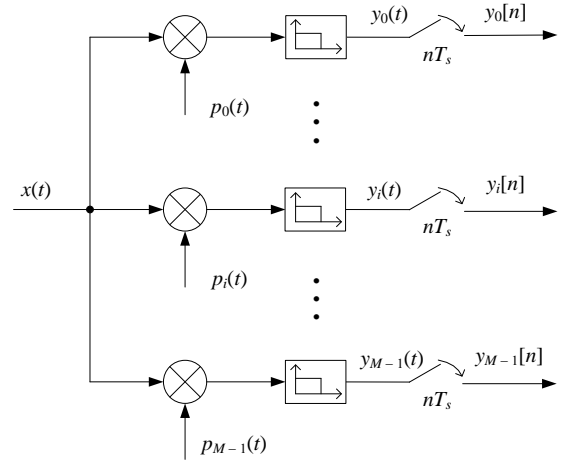

Fig. 5. Block diagram of MWC.

*Modulated wideband converter (MWC)* [17]: MWC is a sub-Nyquist sampling model based on random demodulation technique, as shown in Fig. 5. Signal is fed into a bank of modulators, and it is modulated by a group of high rate pseudorandom sequences, which smear the signal spectrum across the entire spectrums. Then, the modulated signals are low pass filtered and uniformly sampled using a bank of ADCs clocked at low rate. Finally, the signal is reconstructed using orthogonal matching pursuit (OMP) [28], [29] based algorithm (In [17], it is also called continuous-to-finite algorithm).

MWC system can be used to sample the sparse multiband signal at sub-Nyquist rate. Similar to MCS, MWC requires multiple sampling channels to stably reconstruct signal, and roughly $N$ sampling channels need to be designed,

$$N \approx 8K \log_2(Q/4K), \quad (24)$$

where $K$ is the number of pair of active signal sub-bands, and $Q$ is the number of spectral slices. Moreover, The pseudorandom sequence should be clocked at the Nyquist rate. Implementation of such a pseudorandom sequence generator would be nontrivial and costly.

TABLE I
COMPARISON BETWEEN SCS, MCS AND MWC

| Sampling approach | MCS | MWC | SCS |
|---|---|---|---|
| Work for sparse multiband signal | Yes | Yes | Yes |
| Power consumption | More | More | Less |
| Architecture complexity | Complex | Complex | Simple |
| Implementation area | Large | Large | Small |
| Average sampling rate | Low | Low | Low |
| Subject to mismatch of sampling phase | Yes | Yes | Yes |

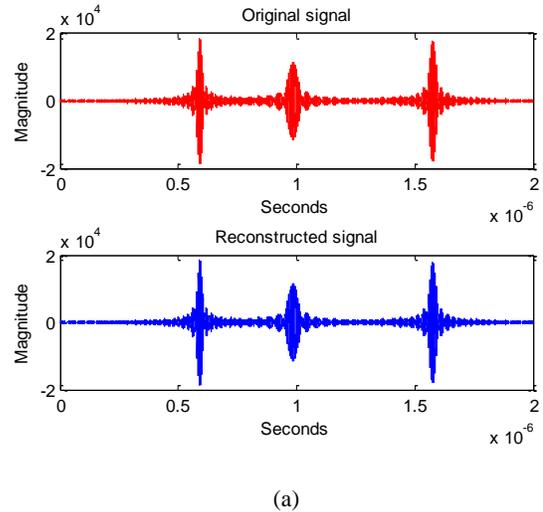

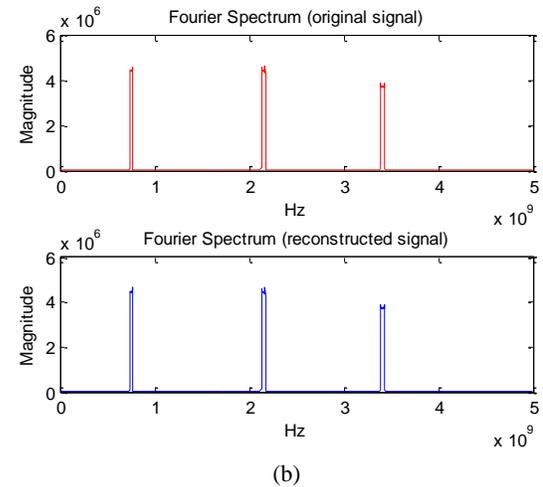

(a)

(b)

Fig. 6. Comparisons of reconstructed waveforms (a) and spectrums (b).

SCS, MCS, and MWC are all based on synchronous sampling, the sampling phase need to be accurately controlled. However, due to non-ideal circuit implementations, the mismatches of offset, gain, and sampling phase among the channels exist. In practical applications, these mismatches will degrade the system performance and should be calibrated. Generally, sampling phase are more difficult to detect and calibrate than the offset and gain mismatches. For SCS and MCS, they can be treated as special cases of time interleaved sampling (TIS) technique (SCS and MCS have fewer channels than TIS). There are many works have been proposed to calibrate the sampling phase [30-32], and the sampling phase calibration is beyond the scope of our work.

Table I presents the comparison between SCS, MCS and MWC. Obviously, SCS is more amenable to implementation. Although SCS and MCS may be subject to ADC bandwidth, the track-and-hold can be adopted to meet this challenge.

## IV. NUMERICAL EXPERIMENTS

In this section, Several numerical experiments are performed to investigate the proposed SCS system.

### A. Setup

An interesting application of SCS is sparse multiband signal, whose band locations are unknown a *priori*. In the experiments, the test signal is defined as

$$x(t) = \sum_{i=1}^{K} \sqrt{E_i B_i} \cdot \text{sinc}(B_i(t-t_i))\cos(2\pi f_i(t-t_i)), \quad (25)$$

where $K$ is the number of active pairs of bands, $E_i$ is the energy coefficient, $B_i$ is the bandwidth of the $i^{th}$ sub-band, $t_i$ is the time offset with respect to $t = 0$, and $f_i$ is the carrier frequency. In all experiments, $t_i$ is randomly chosen in [1, 10] $n$s, $B_i$ is randomly chosen in [20, 50] MHz, and $E_i$ is randomly chosen in [1, 10].

The equivalent sampling rate of reconstructed signal is $f_{Nyq} = $ 10 GHz. The values of co-prime numbers $L_1$ and $L_2$ are set as 6 and 17, then the sampling rates of SCS ADCs are $f_1 = f_{Nyq}/6$ and $f_2 = f_{Nyq}/17$ respectively. The average sampling rate is $f_{avg} \approx$ 2.25 GHz, which is much lower than $f_{Nyq}$.

According to the above setting, signal spectrum is partitioned into $L = 102$ spectral slices with bandwidth of 98 MHz. In order to limit signal bandwidth to $f_{Nyq}$, we require that $f_i + B_i/2 \leq f_{Nyq}/2$. Therefore, $f_i$ is randomly chosen in $[B_i/2, (f_{Nyq} - B_i)/2]$. Since $f_i$ is randomly chosen, the sub-band may be divided into two spectral slices. For real-valued signal $x(t)$ with $K = 3$ active pairs of bands, there would be at most 12 non-zero entries in $\mathbf{x}(f)$.

Signal-to-noise ratio (SNR) defined below is used as a metric to compare the quality of the reconstructed analog waveform:

$$SNR = 20 \cdot \log_{10}\left(\frac{\|\mathbf{x}\|_2}{\|\mathbf{x} - \mathbf{x}^{\#}\|_2}\right). \quad (26)$$

Where $\mathbf{x}$ is the original signal, and $\mathbf{x}^{\#}$ is the reconstructed signal vector.

### B. Experimental Results

In this experiment, we investigate the feasibility of proposed sampling technique. The test signal is noise-free, which is uniformly sampled using SCS system. Samples are re-sequenced based on (5) and (6). Signal is reconstructed from only 10 sampling sequences. Reconstructed waveforms and spectrums are shown in Fig. 6(a) and Fig. 6(b) respectively. The reconstructed signal achieves an SNR = 56.8 dB. Obviously, the feasibility of SCS is demonstrated.

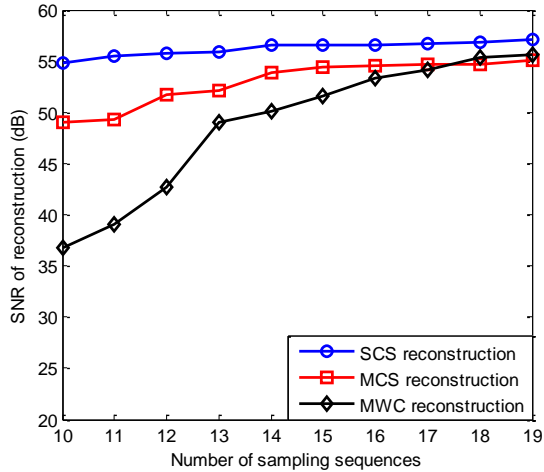

Fig. 7. Reconstruction with different numbers of sampling sequences.

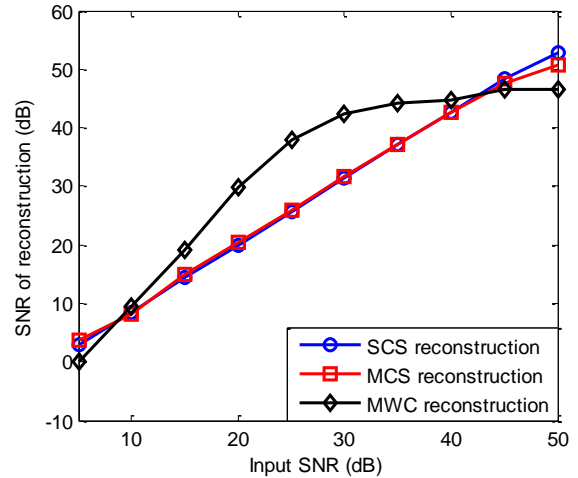

Fig. 8. Reconstruction with different SNRs of input signal.

Generally, more sampling sequences can produce a reconstruction with higher accuracy. In this experiment, we consider the reconstruction performance with respect to the number of sampling sequences that are used in the reconstruction stage. Since both MCS and MWC are applicable to sparse multiband signal, we evaluate their reconstructions. For range of 10 to 19 sampling sequences in increment of 1 sequence are tested, 200 random trials are performed for each specific number. Average SNR of reconstruction for noise-free signal with different numbers of sampling sequences is depicted in Fig. 7. Clearly, SCS outperforms MCS and MWC. Although the principles of SCS and MCS are similar, the patterns of selected sampling sequences of SCS may be more evenly distributed in the block. Therefore, the reconstruction of SCS is with higher accuracy. MWC is based on CS theory, and it can compress signal in the sampling stage. However, more sampling sequences are required to reconstruct signal with a desired accuracy. It is clear that after the number of sampling sequences increases beyond 16, MWC reconstruction yields SNRs close to MCS and SCS reconstructions. On the other hand, for MCS and MWC models, number of sequences means the number of samplers, and system architectures are complex. However, SCS only requires two samplers, and its potential hardware implementation is simple. The advantage of SCS is demonstrated.

In practical application, the signal may be corrupted by noise, or the noise may be introduced in the sampling stage. We consider more practical situation that the white Gaussian noise is added in the test signal, and the noise energy is scaled so that the test signal has a desired SNR. The input signal with SNR over the range of 5 to 50 dB in increment of 5 dB are tested. 200 random trials are performed for each specific SNR value. Signal is reconstructed from 14 sampling sequences, and the average SNR of reconstruction is shown in Fig. 8. Clearly, the proposed SCS is robust against additive white Gaussian noise. MCS and SCS have similar performance. When the input SNRs are within the range [10, 40] dB, MWC outperforms MCS and SCS.

## V. CONCLUSION

In this paper, an acquisition receiver based on synchronous co-prime sampling technique is proposed. In the proposed SCS system, only two samplers are required, which are synchronously and uniformly clocked at sub-Nyquist sampling rates. Specially, the two sampling rates are relatively co-prime. Two samplers digitize a common signal, and SCS only performs one acquisition before reconstruction. The obtained samples are re-sequenced to a group of low speed uniform sequences. We have derived an explicit equation between SCS sampling sequences and unknown signal in the frequency domain. Compared to the conventional sub-Nyquist sampling techniques, such as MCS and MWC, the proposed SCS model not only has the benefits of low sampling rate, but also is with simple system architecture. The proposed approach has been tested using synthetic sparse multiband signal with satisfactory results.

**Appendix A**

Rank Analysis for $\mathbf{R}_y$:

Consider rank($\mathbf{R}_y$), we have following expression

$$rank(\mathbf{R}_y) = rank(\mathbf{\Phi}_S [\mathbf{R}_x]_S [\mathbf{\Phi}_S]^H) \\ \leq \min(rank(\mathbf{\Phi}_S), rank([\mathbf{R}_x]_S [\mathbf{\Phi}_S]^H)). \quad (27)$$

Similarly,

$$rank([\mathbf{R}_x]_S [\mathbf{\Phi}_S]^H) \leq \min(rank([\mathbf{R}_x]_S), rank([\mathbf{\Phi}_S]^H)). \quad (28)$$

Because $rank\left(\left[\mathbf{\Phi}_S\right]\right) = |S|$ and $rank\left(\left[\mathbf{R_x}\right]_S\right) = |S|$, we have

$$rank\left(\mathbf{R_y}\right) \leq |S|. \tag{29}$$

According to Frobenius inequation,

$$\begin{aligned}rank\left(\mathbf{R_y}\right) &= rank\left(\mathbf{\Phi}_S\left[\mathbf{R_x}\right]_S\left[\mathbf{\Phi}_S\right]^H\right) \\ &\geq rank\left(\mathbf{\Phi}_S\left[\mathbf{R_x}\right]_S\right) + rank\left(\left[\mathbf{R_x}\right]_S\left[\mathbf{\Phi}_S\right]^H\right) - rank\left(\left[\mathbf{R_x}\right]_S\right)\end{aligned}. \tag{30}$$

On the other hand,

$$\begin{aligned}rank\left(\mathbf{\Phi}_S\left[\mathbf{R_x}\right]_S\right) &= rank\left(\mathbf{\Phi}_S\mathbf{I}_{|S|}\left[\mathbf{R_x}\right]_S\right) \\ &\geq rank\left(\mathbf{\Phi}_S\mathbf{I}_{|S|}\right) + rank\left(\mathbf{I}_{|S|}\left[\mathbf{R_x}\right]_S\right) - rank\left(\mathbf{I}_{|S|}\right) \\ &= |S|\end{aligned} \tag{31}$$

and

$$\begin{aligned}rank\left(\left[\mathbf{R_x}\right]_S\left[\mathbf{\Phi}_S\right]^H\right) &= rank\left(\left[\mathbf{R_x}\right]_S\mathbf{I}_{|S|}\left[\mathbf{\Phi}_S\right]^H\right) \\ &\geq rank\left(\left[\mathbf{R_x}\right]_S\mathbf{I}_{|S|}\right) + rank\left(\mathbf{I}_{|S|}\left[\mathbf{\Phi}_S\right]^H\right) - rank\left(\mathbf{I}_{|S|}\right) \\ &= |S|\end{aligned} \tag{32}$$

where $\mathbf{I}_{|S|} \in R^{|S| \times |S|}$ is the identity matrix. Therefore, (30) can be rewritten as

$$rank\left(\mathbf{R_y}\right) \geq |S|. \tag{33}$$

Combine (29) and (33), we can obtain that $rank\left(\mathbf{R_y}\right) = |S|$.